\documentclass[
 amsmath,amssymb,aps,twocolumn,a4paper,
]{revtex4-1}

\usepackage{graphicx}
\usepackage{dcolumn}
\usepackage{bm}
\usepackage{lipsum}

\newcommand{\mv}[1]{\ensuremath{\mathbf{#1}}} 
\newcommand{\gv}[1]{\ensuremath{\mbox{\boldmath$ #1 $}}} 
\newcommand{\abs}[1]{\left| #1 \right|} 
\newcommand{\avg}[1]{\left \langle #1 \right \rangle} 
\newcommand{\der}[2]{\frac{d #1}{d #2}} 
\newcommand{\pd}[2]{\frac{\partial #1}{\partial #2}} 

\usepackage[colorlinks]{hyperref}  
\usepackage{cleveref}
\hypersetup{linkcolor=blue,citecolor=blue,urlcolor=blue}

\usepackage{titlesec}
\titleformat{\section}
  {\centering\normalfont\fontsize{11}{15}\bfseries}{\thesection}{1em}{}

\usepackage[toc]{appendix}

\crefname{figure}{fig.}{figs.}
\Crefname{figure}{Figures}{Figures}

\begin{document}

\setcounter{page}{1} 

\title{Active Glassy Dynamics is Unaffected by the Microscopic Details of Self-Propulsion}

\author{Vincent E. Debets$^{1,2}$, Liesbeth M.C. Janssen$^{1,2*}$}

\affiliation{$^{1}$Department of Applied Physics, Eindhoven University of Technology, P.O. Box 513,
5600 MB Eindhoven, The Netherlands\\$^{2}$Institute for Complex
Molecular Systems, Eindhoven University of Technology, P.O. Box 513,
5600 MB Eindhoven, The Netherlands\\}
\email{l.m.c.janssen@tue.nl}

\begin{abstract}%
{\noindent 
Recent years have seen a rapid increase of interest in dense active materials, which, in the disordered state, share striking similarities with conventional passive glass-forming matter.
For such passive glassy materials, it is well established (at least in three dimensions) that the details of the microscopic dynamics, e.g., Newtonian or Brownian, do not influence the long-time glassy behavior. Here we investigate whether this still holds true in the non-equilibrium active case by considering two simple and widely used active particle models, i.e., active Ornstein-Uhlenbeck particles (AOUPs) and active Brownian particles (ABPs). In particular, we seek to gain more insight into the role of the self-propulsion mechanism on the glassy dynamics by deriving a mode-coupling theory (MCT) for thermal AOUPs, which can 
be directly compared to a recently developed MCT for ABPs. Both theories explicitly take into account the active degrees of freedom. We solve the AOUP- and ABP-MCT equations in two dimensions and demonstrate that both models give almost identical results for the intermediate scattering function over a large variety of control parameters (packing fractions, active speeds, and persistence times). We also confirm this theoretical equivalence between the different self-propulsion mechanisms numerically via simulations of a polydisperse mixture of active quasi-hard spheres, thereby establishing that, at least for these model systems, the microscopic details of self-propulsion do not alter the active glassy behavior.}
\end{abstract}

\maketitle 

\section*{Introduction}
\noindent The study of active matter has been gaining widespread attention in the field of colloidal and biological physics since the beginning of the previous decade~\cite{Bechinger2016,Ramaswamy2010,Marchetti2013rev}. While much focus has already been dedicated to dilute and moderately dense self-propelled particle suspensions, recent years have also seen a rising interest in high-density active materials~\cite{Janssen2019active,Berthier2019review}. Interestingly, when self-propelled particles are pushed to sufficiently high densities, regardless of their intrinsic driving, they will manifestly become kinetically arrested and in fact show a strong resemblance to more conventional passive glassy materials. This so-called active glassy behavior has indeed been reported in the context of, e.g.,  living cells~\cite{Zhou2009cell,Parry2014bacterial,Angelini2011cell,Nishizawa2017cell,Garcia2015cell,Grosser2021,Lama2022} and colloidal and granular experiments~\cite{Klongvessa2019colloid1,Klongvessa2019colloid2,Arora2022}, while it has also been observed in multiple theoretical and simulation studies~\cite{Voigtmann2017,SzamelABP2019,SzamelAOUP2015,SzamelAOUP2016,FengHou2017,BerthierABP2014,DijkstraABP2013,BerthierAOUP2017,Berthier2013activeglass,Flenner2020,FlennerAOUP2016,Henkes2011active,Reichert2020modecoupling,Reichert2020tracer,Reichert2021rev,Nandi2018,Sollich2020,janzen2021aging,Janssen2017aging,Bi2016cell,Debets2021cage,paoluzzi2022,Keta2022}.       
Intuitively, it might be expected that dense active matter will eventually be dominated by interactions. However, activity can certainly influence glassy materials in non-trivial ways~\cite{DijkstraABP2013,Berthier2013activeglass,BerthierABP2014,SzamelAOUP2015,FlennerAOUP2016,Flenner2020,Debets2021cage,Keta2022}. The question of what this influence precisely encompasses, and to what degree it depends on the specific details of the active self-propulsion mechanism, has therefore unfolded itself as an increasingly interesting new area of research.   

Two of the simplest and most widely studied models in (dense) active matter are so-called active Brownian particles (ABPs) and active Ornstein Uhlenbeck particles (AOUPs). Their difference rests in the implementation of the self-propulsion force, which either has a constant magnitude and undergoes rotational diffusion (ABPs), or evolves in time according to an Ornstein-Uhlenbeck process (AOUPs). This difference, however, is washed out on a coarse-grained level where the active degrees of freedom are integrated out, in which case both models become identical~\cite{FengHou2017}. Since most theoretical attempts to study dense assemblies of these model active particles have required coarse-graining~\cite{SzamelABP2019,SzamelAOUP2016,FengHou2017,Farage2015,SzamelAOUP2015,Nandi2017MCT}, it has not yet been possible to pinpoint the effect of the specific self-propulsion mechanism on the glassy dynamics.

An exception to the coarse-grained strategy is recent work where a mode-coupling theory (MCT) for ABPs has been developed in which the active degrees of freedom, i.e., the orientations of the active force, are explicitly taken into account~\cite{Voigtmann2017,Reichert2020modecoupling,Reichert2020tracer,Reichert2021rev}. This has revealed several non-trivial short-time features which cannot be captured when employing coarse-grained approaches. A key question, however, remains whether these microscopic details play a significant role in the long-time glassy dynamics of dense active matter. For passive systems it is well-confirmed that (at least in three-dimensional systems) both Brownian and Newtonian dynamics yield identical long-time behavior, and hence the microscopic details of motion are irrelevant for the glassy dynamics. This has been demonstrated both in theory and simulations~\cite{ciarella2021,Flenner2005,Nauroth1997,Sciortino2001,Debets2021Brownian}. It would be interesting to see if such an equivalence is maintained for active systems. Since particle motion becomes more impeded by repulsion at high densities, one would expect the precise single-particle dynamics, whether active or passive, to become increasingly less relevant.

Here we shed more light on the influence of the self-propulsion mechanism on active glassy dynamics from a theoretical perspective. We provide, for the first time, a detailed derivation of a mode-coupling theory for thermal AOUPs which explicitly takes into account the active degrees of freedom. Our theory, which is based on similar principles as the recently developed MCT for ABPs~\cite{Voigtmann2017,Reichert2020modecoupling,Reichert2020tracer,Reichert2021rev}, thus allows for a convenient comparison between both models in the high-density regime. We numerically solve the relevant equations and show that for a wide variety of different settings (packing fractions, active speeds, and persistence times) ABPs and AOUPs give almost identical results after mapping their single-particle dynamics onto each other. To further verify the equivalence between both active self-propulsion models, we also directly compare our theoretical results to ones obtained from simulations of a polydisperse mixture of self-propelled quasi-hard spheres.  

\section*{Theory}
\subsection*{Active Particle Models}
\noindent Both the ABP and AOUP model describe a two-dimensional (2D) $N$-particle active fluid of area $V$ (and number density $\rho=N/V$) as a collection of self-propelling and interacting particles. In particular, each particle $i$ within the fluid evolves in time $t$ according to \cite{Bechinger2016,Farage2015,DijkstraABP2013,FengHou2017}
\begin{equation}\label{eom_r}
    \der{\mv{r}_{i}}{t} = \zeta^{-1} \left( \mv{F}_{i} + \mv{f}_{i} \right) + \gv{\xi}_{i}.
\end{equation}
Here, $\mv{r}_{i}$ denotes the position of particle $i$, $\zeta$ the friction constant, $\mv{F}_{i}$ and $\mv{f}_{i}$ the interaction and self-propulsion force acting on particle $i$ respectively, and $\gv{\xi}_{i}$ a Gaussian thermal noise with zero mean and variance $\avg{\gv{\xi}_{i}(t)\gv{\xi}_{j}(t^{\prime})}_{\mathrm{noise}}=2D\mv{I}\delta_{ij}\delta(t-t^{\prime})$, with $D$ the thermal diffusion coefficient and $\mv{I}$ the unit matrix. 
The distinction between both models resides in the dynamics of the self-propulsion force $\mv{f}_{i}$. For  AOUPs, the time evolution of the self-propulsion force is governed by an Ornstein-Uhlenbeck process~\cite{SzamelAOUP2016,FlennerAOUP2016,SzamelAOUP2015,BerthierAOUP2017,Flenner2020,FengHou2017}
\begin{equation}
    \der{\mv{f}_{i}}{t} = -\tau^{-1} \mv{f}_{i} + \gv{\eta}_{i},
\end{equation} 
where $\tau$ depicts the typical decay time of the self-propulsion and $\gv{\eta}_{i}$ an internal Gaussian noise process with zero mean and a variance $\avg{\gv{\eta}_{i}(t)\gv{\eta}_{j}(t^{\prime})}_{\mathrm{noise}}=2D_{f}\mv{I}\delta_{ij}\delta(t-t^{\prime})$ whose amplitude is controlled by the noise strength $D_{f}$. In contrast, the ABP model assumes a constant absolute value of the self-propulsion speed $v_{0}$, so that $\zeta^{-1}\mv{f}_{i}=v_{0}\mv{e}_{i}=v_{0} [\cos(\theta_{i}),\sin(\theta_{i})]$, and lets the orientation angles $\theta_{i}$ undergo rotational diffusion with a diffusion coefficient $D_{r}$. This yields~\cite{Bechinger2016,Voigtmann2017,SzamelABP2019}
\begin{equation}
    \dot{\theta}_{i} = \chi_{i},
\end{equation}
with $\chi_{i}$ a Gaussian noise process with zero mean and variance $\avg{\chi_{i}(t)\chi_{j}(t^{\prime})}_{\mathrm{noise}}=2D_{\mathrm{r}}\delta_{ij}\delta(t-t^{\prime})$. 

Without particle-particle interactions, both models predict a persistent random walk (PRW), which implies that the mean square displacement (MSD) of each particle is given by~\cite{FengHou2017}
\begin{equation}\label{MSDsingle}
    \avg{\delta r^{2}(t)}=4Dt + 2v_{\mathrm{a}}^{2} \tau_{\mathrm{p}} \left(\tau_{\mathrm{p}}(e^{-t/\tau_{\mathrm{p}}} - 1) + t \right).
\end{equation}
The parameters describing such a PRW are the persistence time, $\tau_{\mathrm{p}}=\tau$ (AOUP), $\tau_{\mathrm{p}}=(D_{\mathrm{r}})^{-1}$ (ABP), an (average) active speed $v_{\mathrm{a}}=v_{0}$ (ABP), $v_{\mathrm{a}}=\sqrt{2D_{f}\tau_{\mathrm{p}}}\zeta^{-1}$ (AOUP), and the thermal diffusion coefficient $D$. On the single-particle level both models can thus strictly be mapped onto each other via the equivalency of their MSDs.

\subsection*{Mode-Coupling Theory}
\noindent To infer information on the collective level, we require the joint $N$-particle probability distribution of positions and self-propulsion forces/orientation angles $P_{N}(\Gamma;t)$. This distribution is governed by the equation
\begin{equation}
\pd{}{t}P_{N}(\Gamma;t)= \Omega P_{N}(\Gamma;t),
\end{equation}
with $\Gamma=(\Gamma_{\mathrm{T}},\Gamma_{\mathrm{R}})=(\mv{r}_{1},\hdots, \mv{r}_{N},\mv{f}_{1},\hdots, \mv{f}_{N})$ (AOUP), $\Gamma=(\Gamma_{\mathrm{T}},\Gamma_{\mathrm{R}})=(\mv{r}_{1},\hdots, \mv{r}_{N},\theta_{1},\hdots, \theta_{N})$ (ABP) denoting the configuration space, and $\Omega$ the evolution operator (see Refs.~\cite{Voigtmann2017,SzamelAOUP2016} for detailed definitions of the latter). Now we assume that our systems can reach a steady-state characterized by a probability distribution $P^{\mathrm{ss}}_{N}(\Gamma)$ that obeys~\cite{SzamelAOUP2016,SzamelABP2019}
\begin{equation}
\Omega P^{\mathrm{ss}}_{N}(\Gamma)=0.
\end{equation}
In principle, we can then study our systems by calculating steady-state averages via
\begin{equation}
\avg{\hdots}=\int d\Gamma \hdots P^{\mathrm{ss}}_{N}(\Gamma).
\end{equation}
However, the steady-state distribution is typically not known exactly. To proceed and make calculations tractable, we will therefore approximate our steady-state averages according to
\begin{equation}
\avg{\hdots}\approx\int d\Gamma \hdots P_{\mathrm{eq}}(\Gamma_{\mathrm{T}})P(\Gamma_{\mathrm{R}}),
\end{equation}
where, for the AOUP model,
\begin{equation}
\begin{split}
    P(\Gamma_{\mathrm{R}}) & =\frac{1}{(2\pi D_{f}\tau_{p})^{dN/2}}\exp \left( - \frac{\sum_{i}\mv{f}_{i}^{2}}{2D_{f}\tau_{p}} \right) \\
    & \hspace{-1.5cm} =  \prod_{i=1}^{N}\frac{1}{(2\pi D_{f}\tau_{p})^{d/2}}\exp \left( - \frac{\mv{f}_{i}^{2}}{2D_{f}\tau_{p}} \right) \equiv \prod_{i=1}^{N}p(\mv{f}_{i})
\end{split}
\end{equation}
represents the distribution of self-propulsion forces, which is factorized in independent Gaussian single-particle distributions $p(\mv{f}_{i})$, while for the ABP model it is simply $P(\Gamma_{\mathrm{R}})=(2\pi)^{-N}$. Note that $d=2$ depicts the dimensionality of the system.  
The distribution of particle positions is the same for both models and given by the Boltzmann solution, $P_{\mathrm{eq}}(\Gamma_{\mathrm{T}})\propto \exp(-\beta U(\Gamma_{\mathrm{T}}))$. This distribution depends solely on the total interaction potential $U(\Gamma_{\mathrm{T}})$, which induces the interaction forces $\mv{F}_{i}=-\nabla_{i}U(\Gamma_{\mathrm{T}})$. Moreover, we assume throughout that the Stokes-Einstein equation connects the inverse thermal energy $\beta$ to the friction constant via $\beta D=\zeta$. As a first approximation, we thus calculate averages based on the distribution the system would assume if the influence of the active forces becomes negligibly small; it therefore neglects any correlations between particle velocities and positions~\cite{Garcia2015cell,FlennerAOUP2016,Szamel_2021} (though we have checked in simulations that these remain relatively small due to the presence of thermal noise) and becomes exact in the limit $v_{\mathrm{a}}\rightarrow 0$. Note that in principle this approximation is similar to the lowest order one in the integration-through-transients formalism, which has been employed in previous work on mode-coupling theory for ABPs and colloidal suspensions under shear flow~\cite{Voigtmann2017,Reichert2021rev,Reichert2020modecoupling,Reichert2020tracer,Fuchs2009}. In this formalism one typically uses transient correlation functions defined with the equilibrium average to find exact expressions for transport coefficients. It has for instance been used to calculate macroscopic stresses in colloidal suspensions. 

In standard mode-coupling theory, the starting point to study the glassy dynamics of a system is the set of density modes~\cite{Janssen2018front,gotze2008complex,Das2004}. Since we want to explicitly include active degrees of freedom, these become more complex in active-MCT and are given by~\footnote{For a motivation of the specific form of the density modes we refer to the derivation in appendix A and Ref.~\cite{Voigtmann2017}}
\begin{equation}
\rho_{l}(\mv{k})= \left\{
\begin{array}{ll}
\frac{i^{m+n}}{\sqrt{N}} \sum_{j=1}^{N} e^{i\mv{k}\cdot \mv{r}_{j}}H_{m}(\bar{f}_{j,x})H_{n}(\bar{f}_{j,y}),\  \mathrm{(AOUP)} \\[5pt]
\frac{1}{\sqrt{N}} \sum_{j=1}^{N}e^{i\mv{k}\cdot \mv{r}_{j}}e^{il\theta_{j}},\hspace{2.5cm}  \mathrm{(ABP)} 
\end{array}
\right..
\end{equation}
Here, $H_{m}(x)$ denotes a normalized Hermite polynomial (see \cref{Hermite}) and $\bar{\mv{f}}\equiv \mv{f}/\sqrt{2D_{f}\tau_{p}}$ a dimensionless self-propulsion force. 
For compactness of notation, we have introduced the index $l$ as a general label for both AOUPs and ABPs; for AOUPs it corresponds to the degree of the Hermite polynomials $l=\{m,n\}$ with $m,n\in[0,\infty ]$, whereas for ABPs it corresponds to the angular mode $l\in[-\infty,\infty]$. 
The equilibrium-averaged (also called transient) time-correlation between such density modes can then be defined via 
\begin{equation}
    S_{l;l^{\prime}}(\mv{k},t)=\avg{\rho^{*}_{l} (\mv{k}) e^{\Omega^{\dagger} t} \rho_{l^{\prime}} (\mv{k})},
\end{equation}
with $\Omega^{\dagger}$ the adjoint evolution operator (see~\cref{Omega_adj} and Ref.~\cite{Voigtmann2017}), which works on everything to its right except for the probability distribution. Note that the lowest order term $S_{0;0}(\mv{k},t)\equiv F(k,t)$ is the same for both models and corresponds to the (transient) intermediate scattering function. It will therefore serve as the main probe to study glassy dynamics of our active systems. Moreover, at time zero, assuming our systems to be isotropic, the density correlation functions are easily calculated and yield
\begin{equation}
\begin{split}
    S_{l;l^{\prime}}(k) & =\avg{\rho^{*}_{l} (\mv{k}) \rho_{l^{\prime}} (\mv{k})} = \delta_{ll^{\prime}}\Big[1+\delta_{l0} (S(k)-1) \Big],
\end{split}
\end{equation}
where $S(k)$ denotes the equilibrium static structure factor, which, for instance, can be obtained from liquid state theory or simulations.

We now follow the mode-coupling strategy pioneered for ABPs in Ref.~\cite{Voigtmann2017} and apply it to AOUPs. The full AOUP MCT derivation is detailed in appendix A. We finally arrive at
the following general equation of motion for the dynamic density correlation functions of both models: 
\begin{equation}\label{eomSt}
\begin{split}
    & \pd{}{t}S_{l ; l^{\prime}}(\mv{k},t) + \sum_{l_{1}} \omega_{l;l_{1}}(\mv{k}) S^{-1}_{l_{1};l_{1}}(k) S_{l_{1} ; l^{\prime}}(\mv{k},t) \\
    & \hspace{0.5cm} + \int_{0}^{t} dt^{\prime} \sum_{l_{1}l_{2}}  M_{l ; l_{1}}(\mv{k},t-t^{\prime}) [\omega^{\mathrm{T}}_{l_{1};l_{2}}(\mv{k})]^{-1} \\
    & \hspace{0.5cm} \times \left[\pd{}{t^{\prime}}S_{l_{2} ; l^{\prime}}(\mv{k},t^{\prime}) + \omega^{\mathrm{R}}_{l_{2};l_{2}}S_{l_{2} ; l^{\prime}}(\mv{k},t^{\prime})\right]=0,
\end{split}
\end{equation}
where $\omega_{l;l^{\prime}}(\mv{k})$ represents the collective diffusion tensor, which governs the short-time dynamics and is split in a translational (T) and rotational (R) term. The memory kernel encodes all non-trivial dynamics and is given by
\begin{equation}\label{Mfinal}
\begin{split}
    M_{l ; l^{\prime}}(\mv{k},t)\approx & \ \frac{\rho}{2} \int \frac{d\mv{q}}{(2\pi)^{2}} \sum_{l_{1}l_{2}}\sum_{l_{3}l_{4}} V_{ll_{1}l_{2}}(\mv{k},\mv{q},\mv{k}-\mv{q}) \\[3pt]
    & \hspace{-1.5cm} \times S_{l_{1} ; l_{3}}(\mv{q},t) S_{l_{2} ; l_{4}}(\mv{k}-\mv{q},t) V^{\mathrm{eq}}_{l^{\prime}l_{3}l_{4}}(\mv{k},\mv{q},\mv{k}-\mv{q}),
\end{split}
\end{equation}
For specific details of the involved parameters, in particular the vertices $V_{ll_{1}l_{2}}(\mv{k},\mv{q},\mv{k}-\mv{q})$ and $V^{\mathrm{eq}}_{ll_{1}l_{2}}(\mv{k},\mv{q},\mv{k}-\mv{q})$, and a precise derivation we refer to Ref.~\cite{Voigtmann2017} and appendix A. We mention that in comparison to the more familiar passive MCT equation~\cite{Nagele1999}, the equation of motion now includes a so-called hopping term $\omega^{\mathrm{R}}_{l_{2};l_{2}}S_{l_{2} ; l^{\prime}}(\mv{k},t^{\prime})$ inside the time integral. This term ensures the long-time decay of the active degrees of freedom~\cite{Voigtmann2017}. Importantly, it must be emphasized that, although the structure of the MCT equation of motion is similar for both models, the individual terms in the equation are not necessarily the same. Most notably the collective diffusion tensor $\omega_{l;l^{\prime}}(\mv{k})$ and the left vertex $V_{ll_{1}l_{2}}(\mv{k},\mv{q},\mv{k}-\mv{q})$ 
harbor the key differences between the AOUP and ABP model. 

To summarize, using only the equilibrium static structure factor $S(k)$, the persistence time $\tau_{\mathrm{p}}$, active speed $v_{\mathrm{a}}$, and area fraction $\phi$ (or number density $\rho$) as input parameters, we can self-consistently find a solution for $S_{l,l^{\prime}}(\mv{k},t)$ and in particular for the intermediate scattering function $F(k,t)$. The latter can then be used to compare the glassy behavior of both models in the high-density regime.

\section*{Methods}
\subsection*{Active-MCT Numerics}
\noindent To establish proof of principle, we numerically solve the active-MCT equations for a monodisperse colloidal mixture of hard disks of diameter $\sigma$. For such a mixture one can employ an analytical expression for $S(k)$ (as a function of the area fraction $\phi=\rho\pi\sigma^{2}$) based on density functional theory~\cite{Thorneywork2018}. The two-dimensional integral over $\mv{q}$ in the memory kernel [\cref{Mfinal}] is rewritten in terms of the coordinates $q=\abs{\mv{q}}$ and $p=\abs{\mv{k}-\mv{q}}$, whose individual integrals are performed on an equidistant wavenumber grid $k\sigma=[0.6,1.0,\hdots,39.8]$. Note that we drop the smallest wavenumber $k\sigma=0.2$ in favor of numerical stability. For computational convenience, we only take into account the lowest order non-trivial active modes, i.e.,\ $l\in [\{0,0\},\{1,0\},\{0,1\}]$ (AOUP) and $l\in [-1,0,1]$ (ABP). It is important to realize that taking the inverse of $\omega^{\mathrm{T}}_{ll^{\prime}}(\mv{k})$ in principle does not commute with the cutoff of active modes. We have checked that taking the inverse at a larger cutoff (up to twenty non-trivial active modes) and afterwards reducing to the lowest order active modes induces slight quantitative changes, but does not qualitatively change our results. Overall, the used cutoff yields stable solutions for the presented range of active speeds and persistence times, although we mention that above the idealized glass transition instabilities on very long time scales still persist. To handle the fact that higher order correlation functions ($S_{l;l^{\prime}}(\mv{k},t)$ with $l,l^{\prime} \neq \{0,0\},0$) depend explicitly on the orientation of the wavevector $\mv{k}$, we can invoke transformation rules that enable us to rewrite correlators with wavevector $\mv{k}$ in terms of ones with a rotated wavevector $\mv{k}^{\prime}$ (see Ref.~\cite{Voigtmann2017} and appendix B for precise details). We can therefore restrict our discussion to wavevectors aligned along a specific direction, which we have chosen to be the $x$-axis, i.e.\ $\mv{k}=k \mv{e}_{x}$. Finally, we fix the passive diffusion coefficient at $D=1$ so that our unit of time equals $\sigma^{2}/D$ and perform the integration over time in~\cref{eomSt} according to the algorithm presented in Ref.~\cite{Voigtmann2017}. For this, we calculate the first $N_{t}/2=16$ points in time using a Taylor expansion with a step size $\Delta t=10^{-6}$, numerically integrate the equations of motion for the next $N_{t}/2$ points in time, duplicate the timestep, and repeat the process.

\subsection*{Simulation Details}
\noindent To complement our theoretical results we also simulate both the AOUP and ABP dynamics of a slightly polydisperse mixture of $N=1000$ quasi-hard disks. Each particle $i$ is described by \cref{eom_r} and the interaction force $\mv{F}_{i}=-\sum_{j \neq i} \nabla_{i} V_{\alpha\beta}(r_{ij})$ is derived from a quasi-hard-sphere powerlaw potential $V_{\alpha\beta}(r)= \epsilon\left( \frac{\sigma_{\alpha\beta}}{r}\right)^{36}$~\cite{Weysser2010structural,Lange2009}. The interaction energy $\epsilon$, friction constant $\zeta$, and diffusion coefficient $D$ are all set to a value of one. To ensure polydispersity, our mixture consists of equal fractions of particles with diameters (in units of $\sigma$) $\sigma_{\alpha\alpha}=\{0.8495,0.9511,1.0,1.0489,1.1505\}$~\footnote{Particle diameters are chosen such that the first four moments correspond to the results of a Gaussian distribution with a mean of $1$ and a standard deviation of $0.1$.}, which are additive so that $\sigma_{\alpha\beta}=(\sigma_{\alpha\alpha}+\sigma_{\beta\beta})/2$. Simulations are performed by solving the Langevin equation (\cref{eom_r}) via a forward Euler scheme and are carried out using LAMMPS~\cite{Lammps}. We fix the square box size to set the area fraction at $\phi=0.75$ (higher values tend to result in  crystallization) 
and impose periodic boundary conditions. We then set the persistence time $\tau_{\mathrm{p}}$ and active speed $v_{\mathrm{a}}$, run the system sufficiently long to ensure no aging takes place, and afterwards track the particles over time. All simulation results are presented in units where $\sigma$, $\epsilon$, and $\zeta \sigma^{2}/\epsilon$ denote the units of length, energy, and time respectively~\cite{Flenner2005}.

\section*{Results \& Discussion}
\noindent Before proceeding to the glassy dynamics, we first briefly discuss the free-particle dynamics in more detail to elucidate potential intrinsic differences between both models. For this we exploit the fact that at zero density the memory kernel can be set to zero and that $S(k)=1$~\cite{Voigtmann2017}. 
\begin{figure}[hb!]
    \centering
    \includegraphics[width=0.5\textwidth]{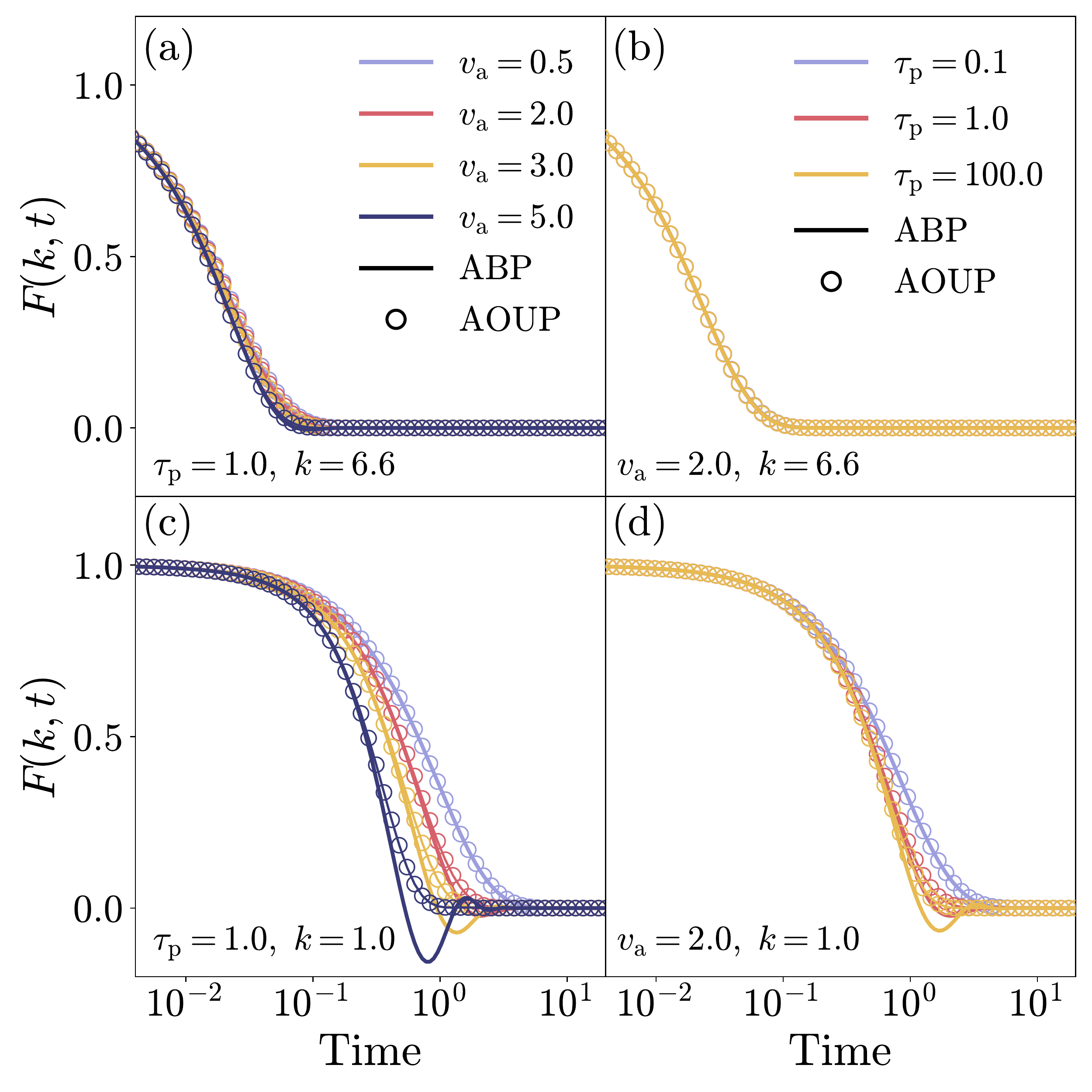}
    \caption{The intermediate scattering function $F(k,t)$ for free particles ($\rho=0$) as a function of time obtained for both ABP-MCT (solid lines) and AOUP-MCT (circles) at (a-b) a wavevector $k=6.6$ close to the first peak of the static structure factor and (c-d) a relatively small wavevector $k=1.0$. Results correspond to (a,c) different active speeds, and (b,d) different persistence times.}
    \label{fig_single}
\end{figure}
This allows us to exactly solve \cref{eomSt} which yields $\mv{S}(\mv{k},t)=\exp(-\gv{\omega}(\mv{k})t)$. 
\begin{figure*}[ht!]
    \centering
    \includegraphics[width=1.0\textwidth]{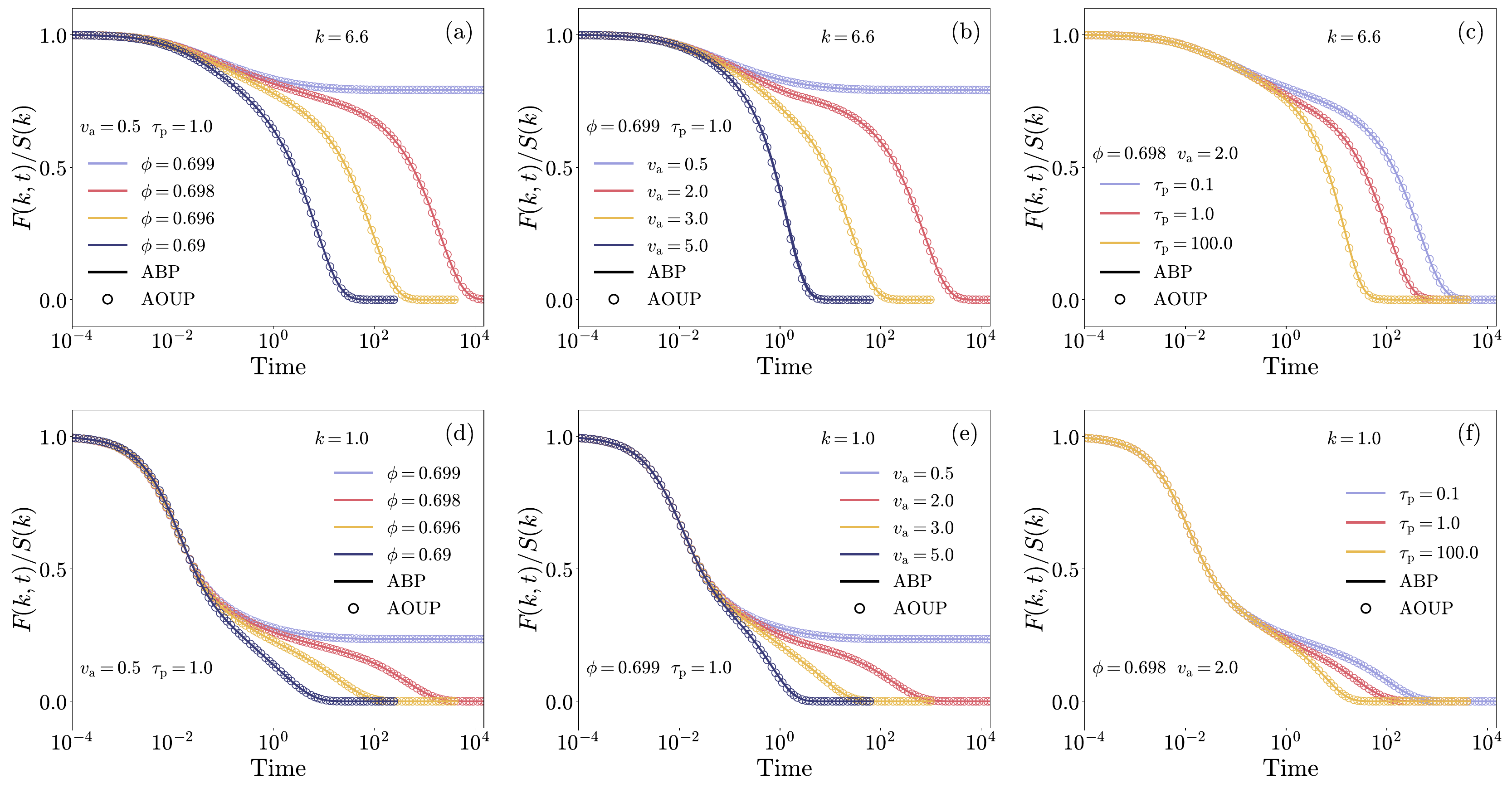}
    \caption{The normalized intermediate scattering function $F(k,t)/S(k)$ as a function of time obtained for both ABP-MCT (solid lines) and AOUP-MCT (circles) at (a-c) a wavevector $k=6.6$ close to the first peak of the static structure factor and (d-f) a relatively small wavevector $k=1.0$. Results correspond to (a,d) different packing fractions, (b,e) different active speeds, and (c,f) different persistence times.}
    \label{figMCT}
\end{figure*}
Based on this result we have calculated the intermediate scattering function $F(k,t)$ for different active speeds and persistence times and have plotted the results in~\cref{fig_single}.
Note that the free-particle solution allows for the inclusion of many active modes, and we have verified that these results remain unaltered upon adding more active modes. An inspection of~\cref{fig_single} shows that $F(k,t)$ decays more rapidly upon increasing the active speed or persistence time. Moreover, at a large wavenumber $k=6.6$ both models give the same results which is consistent with our initial mapping of the single-particle MSDs [see \cref{MSDsingle}]. Interestingly, it can be seen that at a relatively small wavenumber, $k=1.0$, differences between both models start to manifest themselves, especially at larger values of the active speed and persistence time. In particular, the ABP model yields oscillatory behavior which has been attributed to the persistent swimming of the ABPs~\cite{Kurzthaler2016,Kurzthaler2018}. These oscillations are absent for the AOUPs since the Ornstein-Uhlenbeck process is Gaussian and correlation functions therefore should decay monotonically. A mapping based on the MSD, which is essentially a zero-wavenumber limit of the density correlation function, thus misses these differences at finite $k$. In other words, the free-particle intermediate scattering function $F(k,t)$ ($\rho=0$) can distinguish between the ABP and AOUP model.

\noindent Let us now look at the theoretical predictions of the ABP- and AOUP-MCT frameworks at high densities to understand their glassy behavior and see whether the single-particle differences between both models persist in the glassy regime. To compare both models we have primarily focused on the intermediate scattering function $F(k,t)$, which has been plotted for a variety of different settings and both models in \cref{figMCT}. We note that, despite the presence of an active self-propulsion mechanism, both ABP-MCT and AOUP-MCT still predict an idealized glass transition upon increasing the packing fraction (or density). This is characterized by the emergence of a nonzero long-time value for $F(k,t)$. Moreover, we see that increasing the active speed $v_{\mathrm{a}}$ and the persistence time $\tau_{\mathrm{p}}$ always yields faster relaxation dynamics, represented by a more rapid decay to zero of $F(k,t)$. These predictions are all consistent with the previous in-depth study of ABP-MCT and simulations of a polydisperse mixture of self-propelling hard spheres~\cite{Voigtmann2017,DijkstraABP2013}, though we mention that an increase of the persistence time at a fixed effective temperature (instead of the active speed) can also yield non-monotonic behavior~\cite{SzamelAOUP2015}. This reentrant dynamics has already been qualitatively predicted by a recently developed MCT for athermal AOUPs~\cite{SzamelAOUP2015,SzamelAOUP2016} and rationalized in terms of efficient cage exploration~\cite{Debets2021cage}. 

More strikingly, however, we observe that for all shown cases and all considered time scales, both models predict almost identical results. This implies that, at least in the numerically accessible region, the mapping between ABPs and AOUPs based on the single-particle MSDs~[\cref{MSDsingle}] transfers directly to the collective structural relaxation in the high-density regime. Interestingly, for a relatively small wavenumber $k=1.0$ the differences on the single-particle level [see~\cref{fig_single}c-d] have even been washed out in the glassy regime with $F(k,t)$ in all cases decaying monotonically. Since an oscillatory decay of $F(k,t)$ has been attributed to persistent swimming of the ABPs, we expect that this is suppressed by particle-particle interactions at high densities. This in turn forces the models to become more equivalent and give almost identical results.  We have also verified that this equivalence occurs over an even larger parameter range than presented in~\cref{figMCT}. This suggests that, at least for the chosen model systems, the long-time dynamics does not depend on the microscopic details of the active self-propulsion, which is consistent with recent simulations of (a)thermal ABPs and AOUPs where a different parameter regime (larger active speeds and smaller persistence times) has been probed~\cite{Debets2021cage}. An important consequence of this equivalence might reside in the modelling of more complex dense active systems, such as confluent cell layers~\cite{Bi2016cell,Lang2018}. For such systems it is often hard to infer precise details of the microscopic self-propulsion mechanism. 
Our results suggest that these details might be of lesser importance when studying high-density active matter.

\begin{figure}[hb!]
    \centering
    \vspace{0.0cm}
    \includegraphics 
    [width=8.8cm,height=4.4cm] 
    {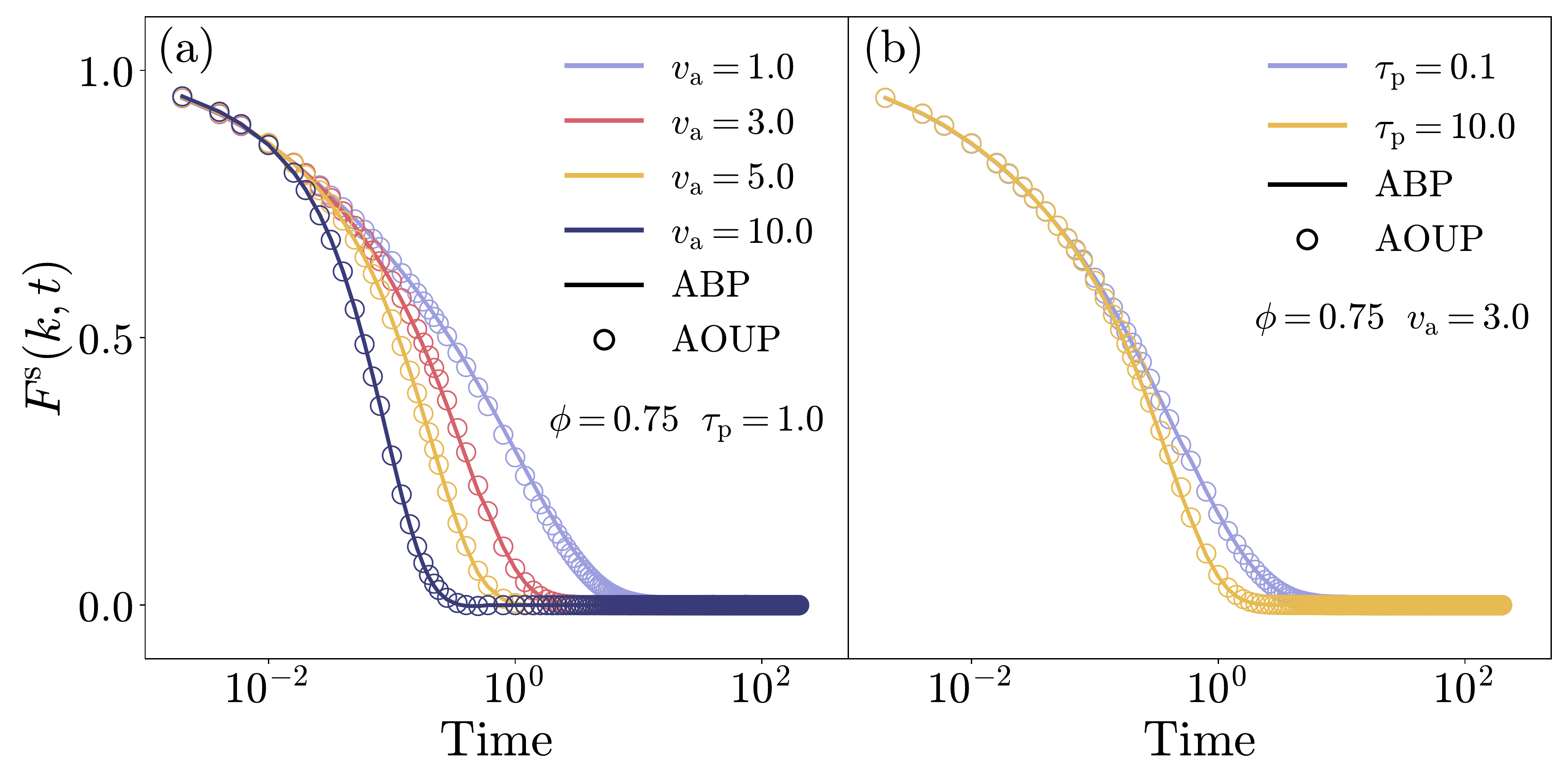} 
    \caption{The self-intermediate scattering function $F^{\mathrm{s}}(k,t)$ as a function of time obtained from both ABP and AOUP simulations at a wavevector $k=6.6$ close to the first peak of the static structure factor. Results correspond to different (a) active speeds, and (b) persistence times.}
    \label{Fig2}
\end{figure}

To place our theoretical findings in a broader context we now proceed to the predictions from our simulations. Based on the retrieved particle trajectories we have calculated the self-intermediate scattering function, i.e., $F^{\mathrm{s}}(k,t)=\avg{e^{-i\mv{k}\cdot \mv{r}_{j}(0)}e^{i\mv{k}\cdot \mv{r}_{j}(t)}}$, where we mention that in simulations the statistical averaging is done with respect to the active steady-state. However, at high densities, the differences between steady-state and transient self-intermediate scattering functions have been found to be small (see~\cite{Reichert2020modecoupling} for a more detailed discussion). The results for both models are plotted for a variety of settings in~\cref{Fig2}. It can be seen that the relaxation of the self-intermediate scattering function occurs on shorter timescales upon increasing the active speed $v_{\mathrm{a}}$ (\cref{Fig2}a) or the persistence time $\tau_{\mathrm{p}}$ (\cref{Fig2}b). These results are qualitatively consistent with our theoretical predictions for the intermediate scattering function and imply that enhanced particle speed and persistence render the material more liquid-like. 

Interestingly, we find that, also for our simulation results, the differences between both model systems are manifestly only marginal. This further substantiates our theoretical predictions and indicates that for simple model active systems the active glassy dynamics is unaffected by the microscopic details of active self-propulsion. This behavior is analogous to more conventional passive glass-forming materials, where it is well established that, at least in three dimensions, different single-particle dynamics, e.g., Newtonian or Brownian, yield similar long-time dynamics~\cite{ciarella2021,Flenner2005,Nauroth1997,Sciortino2001}. 

\begin{figure}[ht!]
    \centering
    \vspace{0.0cm}
    \includegraphics 
    [width=8.8cm,height=4.4cm] 
    {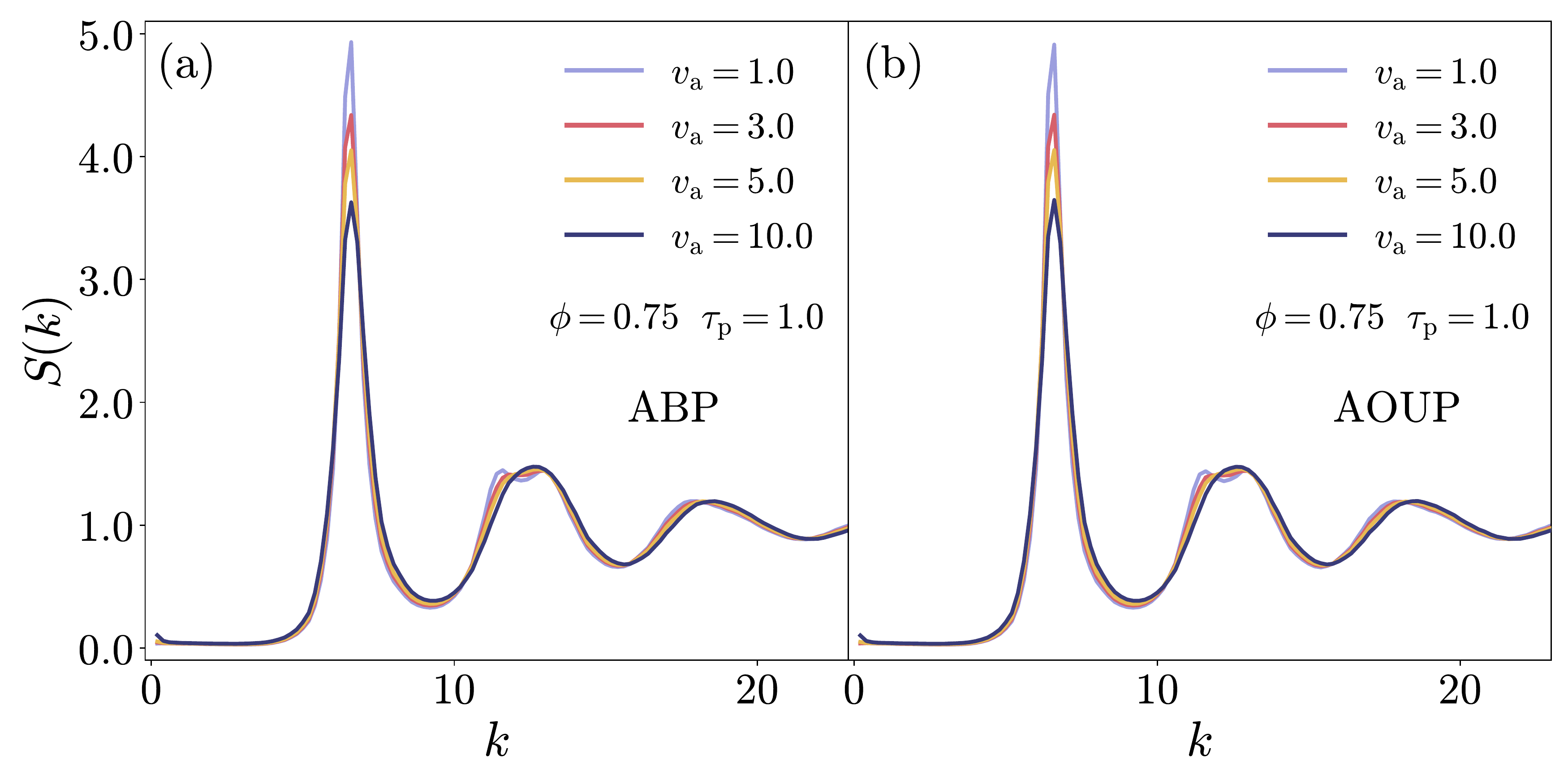} 
    \caption{The static structure factor $S(k)$ as a function of the wavevector $k$ obtained from both (a) ABP and (b) AOUP simulations. Results correspond to different active speeds.}
    \label{Fig3}
\end{figure}

We finalize our discussion by mentioning that a critical assumption in the presented active-MCT theories is the replacement of the steady-state probability distribution by its equilibrium counterpart. In other words, we assume the same (passive) structure for both models. Although unlikely, structural differences between both models might therefore still exist. To verify that our mapping of the dynamics also yields similar structures, we have retrieved the static structure factor, i.e., $S(k)=\avg{\sum_{j=1}^{N}e^{-i\mv{k}\cdot \mv{r}_{j}}\sum_{l=1}^{N}e^{i\mv{k}\cdot \mv{r}_{l}}}$, for the same parameters as for the self-intermediate scattering function. The results for different active speeds are plotted in~\cref{Fig3}. Consistent with the dynamical quantities we see that the static structure factor also remains almost unaltered when we interchange the ABP and AOUP model. Moreover, we see the height of the first peak decreasing upon increasing the active speed of the particles, which is consistent with the faster relaxation dynamics observed for the self-intermediate scattering function.

\section*{Conclusion}
\noindent In this work we have presented the first derivation of an MCT for thermal AOUPs that explicitly takes into account the active degrees of freedom (self-propulsion forces) via the density modes. Our derivation is based on previous work on ABP-MCT and uses the same assumption of replacing steady-state averages by their equilibrium-averaged or transient counterparts. The central result comprises an equation of motion for the (transient) intermediate scattering function, which can be self-consistently solved using only the equilibrium static structure factor and relevant control parameters (packing fraction, active speed, and persistence time) as input. 

Interestingly, after mapping both models on the single-particle (or non-interacting) level via their MSDs, our newly developed AOUP-MCT gives almost identical results as ABP-MCT over a wide range of values for the different control parameters. In other words, the equivalence of both models in the non-interacting regime transfers directly to the collective relaxation in the high-density (glassy) regime. Although this is consistent with recent simulation results~\cite{Debets2021cage}, we have further confirmed the witnessed equivalence between the different self-propulsion mechanisms by performing simulations of a polydisperse mixture of active quasi-hard spheres. In all cases, the differences between the AOUP and ABP simulations are minute. We thus conclude that, at least for the considered model systems, the microscopic details of the self-propulsion do not affect the active glassy behavior.

As a followup it would be intriguing to see whether the witnessed equivalence between both models in the glassy regime can also be formally established given that the structure of the derived MCT equations is already identical. This could provide crucial theoretical insight into the emergent universality of dense active matter. A possible starting point for it might be to try to convert the Hermite-polynomial basis for the AOUPs into the trigonometric one adopted for the ABPs. Alternatively, one can look into the scaling laws close to the idealized glass transition, which have already been extensively studied for passive MCT~\cite{gotze2008complex}.

It could also be interesting to test the validity of the observed equivalence for more complex active self-propulsion models or when transitioning from overdamped to underdamped active dynamics (from microswimmers to so-called microflyers~\cite{Lowen2020active}). Finally, we mention that the derivation of AOUP-MCT can be easily extended to three dimensions. We therefore hope that the framework of AOUP-MCT will continue to be used for comparison with simulation or experimental results in order to better understand the rich phenomenology of active glassy matter.

\section*{Acknowledgments}
\noindent We acknowledge the Dutch Research Council (NWO) for financial support through a START-UP grant (V.E.D. and L.M.C.J.).

\appendix

\section{Derivation of AOUP-MCT}
\label{AppA}

\noindent  As our starting point for the AOUP-MCT derivation, we take the following fluctuating local density to describe the collective motion of 2D AOUPs:
\begin{equation}
\begin{split}
    \rho(\mv{r},\mv{f}) & =\frac{1}{p(\mv{f})}\sum_{i=1}^{N}\delta (\mv{r}-\mv{r}_{i})\delta (\mv{f}-\mv{f}_{i})\\
    &=\pi^{3/2}\exp \left(  \bar{\mv{f}}^{2} \right)\sum_{i=1}^{N}\delta (\mv{r}-\mv{r}_{i})\delta (\bar{\mv{f}}-\bar{\mv{f}}_{i}).
\end{split}
\end{equation}
Here, we have introduced a dimensionless self-propulsion force $\bar{\mv{f}}\equiv \mv{f}/\sqrt{2D_{f}\tau_{p}}$ and added a prefactor for normalization. Next, we Fourier-Hermite expand the microscopic density, i.e.\ 
\begin{equation}
    \rho(\mv{r},\mv{f})= \frac{1}{V}\sum_{\mv{k}}\sum_{mn}(-i)^{m+n}\rho_{mn}(\mv{k})e^{-i\mv{k}\cdot \mv{r}} H_{m}(\bar{f}_{x})H_{n}(\bar{f}_{y}),
\end{equation}
with the factor $(-i)^{m+n}$ added for technical convenience and the normalized Hermite polynomials being defined as
\begin{equation}\label{Hermite}
    H_{n}(x)=\frac{1}{\sqrt{2^{n}n!}}(-1)^{n}e^{x^{2}}\frac{d^{n}}{dx^{n}}e^{-x^{2}}.
\end{equation}
Invoking the orthogonality of the Hermite polynomials with respect to the measure $\exp (-\bar{\mv{f}}^{2})$ we obtain for the density modes
\begin{equation}
    \rho_{mn} (\mv{k})=\frac{1}{\sqrt{N}}i^{m+n}\sum_{j=1}^{N} e^{i\mv{k}\cdot \mv{r}_{j}}H_{m}(\bar{f}_{j,x})H_{n}(\bar{f}_{j,y}).
\end{equation}
The transient (or equilibrium-averaged) time-correlation between such density modes can then be defined via 
\begin{equation}
    S_{mn;m^{\prime}n^{\prime}}(\mv{k},t)=\avg{\rho^{*}_{mn} (\mv{k}) e^{\Omega^{\dagger} t} \rho_{m^{\prime}n^{\prime}} (\mv{k})},
\end{equation}
where the adjoint (or backward) evolution operator is given by
\begin{equation}\label{Omega_adj}
\begin{split}
    \Omega^{\dagger} = \Omega^{\dagger}_{\mathrm{T}} + \Omega^{\dagger}_{\mathrm{R}} & = \sum_{i=1}^{N}  \left( D \nabla_{i} + \zeta^{-1}  (\mv{F}_{i} + \mv{f}_{i}) \right) \cdot \nabla_{i} \\
    & + \sum_{i=1}^{N} \left( \left( D_{f}\pd{}{\mv{f}_{i}} -\tau_{p}^{-1} \mv{f}_{i} \right) \cdot \pd{}{\mv{f}_{i}} \right) .
\end{split}
\end{equation}
We adopt the convention that the adjoint evolution operator works on everything to its right except for the probability distribution. Note that the lowest order term $S_{00;00}(\mv{k},t)\equiv F(k,t)$ corresponds to the intermediate scattering function, which will serve as the main probe to study glassy dynamics of our active system. 
At time zero, assuming our system to be isotropic, the density correlation functions are easily calculated and yield
\begin{equation}
\begin{split}
    S_{mn;m^{\prime}n^{\prime}}(k) & =\avg{\rho^{*}_{mn} (\mv{k}) \rho_{m^{\prime}n^{\prime}} (\mv{k})} \\[3pt]
    & \hspace{-0.0cm} = \delta_{mm^{\prime}}\delta_{nn^{\prime}}\Big[1+\delta_{m0}\delta_{n0} (S(k)-1) \Big],
\end{split}
\end{equation}
where $S(k)$ denotes the equilibrium static structure factor.

To arrive at an equation of motion for the intermediate scattering function we will employ a similar strategy as already introduced for ABPs in Ref.~\cite{Voigtmann2017}. Starting from the Mori-Zwanzig projector formalism~\cite{Mori65,Zwanzig60}, we introduce a projector on density modes (using the shorthand notation $l\equiv \{ m,n\}$ which will be done throughout)
\begin{equation}
\begin{split}
    \mathcal{P} & =\sum_{l_{1}}\sum_{l_{2}} \rho_{l_{1}}(\mv{k})\big\rangle S^{-1}_{l_{1};l_{2}}(k) \big\langle \rho^{*}_{l_{2}}(\mv{k})\\
    & =\sum_{l_{1}} \rho_{l_{1}}(\mv{k})\big\rangle S^{-1}_{l_{1};l_{1}}(k) \big\langle \rho^{*}_{l_{1}}(\mv{k}),
\end{split}
\end{equation}
and its orthogonal counterpart $\mathcal{Q}=1-\mathcal{P}$. Note that the superscript $-1$ represents the inverse matrix of the respective quantity, i.e.\ $X^{-1}_{l;l^{\prime}}\equiv [\mathbf{X}^{-1}]_{l;l^{\prime}}$. Following standard procedure in MCT one can then derive that
\begin{equation}\label{eomF1}
\begin{split}
    \pd{}{t}S_{l ; l^{\prime}}(\mv{k},t) & + \sum_{l_{1}}  \omega_{l;l_{1}}(\mv{k}) S^{-1}_{l_{1};l_{1}}(k) S_{l_{1} ; l^{\prime}}(\mv{k},t) \\
    & \hspace{-2.0cm} - \int_{0}^{t} dt^{\prime}\sum_{l_{1}} K_{l;l_{1}}(\mv{k},t-t^{\prime}) S^{-1}_{l_{1};l_{1}}(k) S_{l_{1} ; l^{\prime}}(\mv{k},t^{\prime}) = 0.
\end{split}
\end{equation}
In this equation the collective diffusion tensor, which governs the short-time dynamics, is given by
\begin{equation}\label{eq_omega}
\begin{split}
    \omega_{l;l^{\prime}}(\mv{k}) & = -\avg{\rho^{*}_{l} (\mv{k}) \Omega^{\dagger} \rho_{l^{\prime}} (\mv{k})} = \Big[ k^{2} D \delta_{ll^{\prime}}  \\[3pt]
    & \hspace{-0.5cm}   + \zeta^{-1} \sqrt{D_{f}\tau_{p}}S_{l;l}(k)\ \mv{k}\cdot (\gv{\delta}_{ll^{\prime -}} - \gv{\delta}_{ll^{\prime +}}) \Big] \\[3pt]
    & \hspace{-0.5cm} + \Big[ (m+n)\tau_{p}^{-1} \delta_{ll^{\prime}} \Big]  \equiv \omega^{\mathrm{T}}_{l;l^{\prime}}(\mv{k}) + \omega^{\mathrm{R}}_{l;l^{\prime}},
\end{split}
\end{equation}
where we have introduced the two-vector ($[\cdot,\cdot]$) shorthand notation
\begin{equation}
\begin{split}
  \gv{\delta}_{ll^{\prime \pm}} & =\bigg[ \sqrt{m^{\prime}+\frac{1}{2}\pm \frac{1}{2}}\ \delta_{m,m^{\prime}\pm1}\delta_{nn^{\prime}}, \\ 
  & \hspace{0.6cm} \sqrt{n^{\prime}+\frac{1}{2}\pm \frac{1}{2}} \delta_{n,n^{\prime}\pm1}\delta_{mm^{\prime}} \bigg].  
 \end{split}
\end{equation}
The memory kernel, which represents all nontrivial dynamics, can be formally written as 
\begin{equation}
\begin{split}
    K_{l ; l^{\prime}}(\mv{k},t) & = \avg{\rho^{*}_{l} (\mv{k}) \Omega^{\dagger} \mathcal{Q} e^{\mathcal{Q}\Omega^{\dagger}\mathcal{Q}t} \mathcal{Q}\Omega^{\dagger} \rho_{l^{\prime}} (\mv{k})}\\[3pt]
    & = \avg{\rho^{*}_{l} (\mv{k}) \Omega^{\dagger}_{\mathrm{T}} \mathcal{Q} e^{\mathcal{Q}\Omega^{\dagger}\mathcal{Q}t} \mathcal{Q}\Omega^{\dagger}_{\mathrm{T}} \rho_{l^{\prime}} (\mv{k})}.
\end{split}
\end{equation}
Here we have used that, since the active degrees of freedom (self-propulsion forces) never slow down, the $\Omega^{\dagger}_{\mathrm{R}}$-terms do not contribute to the vertices, i.e.\ $\mathcal{Q}\Omega^{\dagger}_{\mathrm{R}}\rho_{l}(\mv{k})\big\rangle=\big \langle \rho_{l}(\mv{k})\Omega^{\dagger}_{\mathrm{R}} \mathcal{Q} = 0$. Consequently, only the translational degrees of freedom yield slow dynamics and we therefore seek to convert the memory kernel to an irreducible (friction) memory kernel by means of the operators 
\begin{equation}
    \mathcal{P}^{\prime} = -\sum_{l_{1}l_{2}} \rho_{l_{1}}(\mv{k})\big\rangle [\omega^{\mathrm{T}}_{l_{1};l_{2}}(k)]^{-1} \big\langle \rho^{*}_{l_{2}}(\mv{k})\Omega^{\dagger}_{\mathrm{T}},
\end{equation}
and $\mathcal{Q}^{\prime}=1-\mathcal{P}^{\prime}$. Invoking Dyson decomposition, we may write
\begin{equation}\label{eomM1}
\begin{split}
    K_{l ; l^{\prime}}(\mv{k},t) = & M_{l ; l^{\prime}}(\mv{k},t) - \int_{0}^{t}dt^{\prime} \sum_{l_{1}l_{2}}  M_{l ; l_{1}}(\mv{k},t-t^{\prime}) \\[3pt]
    & \times [\omega^{\mathrm{T}}_{l_{1};l_{2}}(\mv{k})]^{-1} K_{l_{2} ; l^{\prime}}(\mv{k},t^{\prime}),
\end{split}
\end{equation}
with the irreducible memory kernel defined as 
\begin{equation}
    M_{l ; l^{\prime}}(\mv{k},t)= \avg{\rho^{*}_{l} (\mv{k}) \Omega^{\dagger}_{\mathrm{T}} \mathcal{Q} e^{\mathcal{Q}\Omega^{\dagger}\mathcal{Q}^{\prime}\mathcal{Q}t} \mathcal{Q}\Omega^{\dagger}_{\mathrm{T}} \rho_{l^{\prime}} (\mv{k})}.
\end{equation}
Now we can combine \cref{eomF1,eomM1} to arrive at an equation of motion for the intermediate scattering function, which lends itself to mode-coupling-like approximations:
\begin{equation}
\begin{split}
    & \pd{}{t}S_{l ; l^{\prime}}(\mv{k},t) + \sum_{l_{1}} \omega_{l;l_{1}}(\mv{k}) S^{-1}_{l_{1};l_{1}}(k) S_{l_{1} ; l^{\prime}}(\mv{k},t) \\
    & \hspace{0.5cm} + \int_{0}^{t} dt^{\prime} \sum_{l_{1}l_{2}}  M_{l ; l_{1}}(\mv{k},t-t^{\prime}) [\omega^{\mathrm{T}}_{l_{1};l_{2}}(\mv{k})]^{-1} \\
    & \hspace{0.5cm} \times \left[\pd{}{t^{\prime}}S_{l_{2} ; l^{\prime}}(\mv{k},t^{\prime}) + \omega^{\mathrm{R}}_{l_{2};l_{2}}S_{l_{2} ; l^{\prime}}(\mv{k},t^{\prime})\right]=0.
\end{split}
\end{equation}
We mention that this equation is identical in structure to the one obtained for ABPs in Ref.~\cite{Voigtmann2017} and reiterate that, in comparison to the more familiar passive MCT equation, there is an additional hopping term $\omega^{\mathrm{R}}_{l_{2};l_{2}}S_{l_{2} ; l^{\prime}}(\mv{k},t^{\prime})$ inside the time integral. This term ensures the long-time decay of the active degrees of freedom. At the same time, we also emphasize that the individual terms in the equation are not necessarily the same as the ones presented in Ref.~\cite{Voigtmann2017} for ABPs, and these terms will therefore harbor the differences between the AOUP and ABP model.

To proceed and find a solution for the active-MCT equation, we project the fluctuating forces $\mathcal{Q}\Omega^{\dagger}_{\mathrm{T}} \rho_{l^{\prime}} (\mv{k})$ onto density doublets. Specifically, we introduce, assuming Gaussian factorization for higher order static correlations~\cite{Janssen2015a} and making use of the fact that $S^{-1}_{l_{1};l_{2}}(q)$ is diagonal, the projection operator
\begin{equation}
\begin{split}
    \mathcal{P}_{2} = \frac{1}{2}\sum_{\mv{q}_{1}\mv{q}_{2}}\sum_{l_{1}l_{2}} &\  \rho_{l_{1}}(\mv{q}_{1}) \rho_{l_{2}}(\mv{q}_{2})\big\rangle S^{-1}_{l_{1};l_{1}}(q_{1})S^{-1}_{l_{2};l_{2}}(q_{2}) \\
    & \times \big\langle \rho^{*}_{l_{1}}(\mv{q}_{1})\rho^{*}_{l_{2}}(\mv{q}_{2}),
\end{split}
\end{equation}
and use it to approximate
\begin{equation}
\begin{split}
    & M_{l ; l^{\prime}}(\mv{k},t)\approx \avg{\rho^{*}_{l} (\mv{k}) \Omega^{\dagger}_{\mathrm{T}} \mathcal{Q} \mathcal{P}_{2}e^{\mathcal{Q}\Omega^{\dagger}\mathcal{Q}^{\prime}\mathcal{Q}t}\mathcal{P}_{2} \mathcal{Q}\Omega^{\dagger}_{\mathrm{T}} \rho_{l^{\prime}} (\mv{k})} \\[3pt]
    & = \sum_{\mv{q}_{1}\hdots \mv{q}_{4}}\sum_{l_{1}\hdots l_{4}} \avg{\rho^{*}_{l} (\mv{k}) \Omega^{\dagger}_{\mathrm{T}} \mathcal{Q} \rho_{l_{1}}(\mv{q}_{1}) \rho_{l_{2}}(\mv{q}_{2})} S^{-1}_{l_{1};l_{1}}(q_{1}) \\[3pt]
    & S^{-1}_{l_{2};l_{2}}(q_{2}) \avg{\rho^{*}_{l_{1}}(\mv{q}_{1})\rho^{*}_{l_{2}}(\mv{q}_{2}) e^{\mathcal{Q}\Omega^{\dagger}\mathcal{Q}^{\prime}\mathcal{Q}t} \rho_{l_{3}}(\mv{q}_{3}) \rho_{l_{4}}(\mv{q}_{4}) } \\[3pt]
    & S^{-1}_{l_{3};l_{3}}(q_{3})S^{-1}_{l_{4};l_{4}}(q_{4}) \avg{ \rho^{*}_{l_{3}}(\mv{q}_{3}) \rho^{*}_{l_{4}}(\mv{q}_{4}) \mathcal{Q} \Omega^{\dagger}_{\mathrm{T}} \rho^{*}_{l^{\prime}} (\mv{k})} .
\end{split}
\end{equation} 
To make this expression tractable we explicitly calculate both vertices. For convenience, we split the translational evolution operator into a passive and active contribution, i.e.\ $\Omega^{\dagger}_{\mathrm{T}}=\Omega_{\mathrm{eq}}^{\dagger}+\delta\Omega_{\mathrm{T}}^{\dagger}$, with $\delta\Omega_{\mathrm{T}}^{\dagger}=\sum_{i}^{N} \zeta^{-1}\mv{f}_{i} \cdot \nabla_{i}$. Moreover, invoking the following orthogonality relation for Hermite polynomials,
\begin{equation}
\begin{split}
    & \pi^{-1/2} \int_{\infty}^{\infty} dx\ H_{m}(x)H_{n}(x)H_{s}(x) e^{-x^{2}} = \sqrt{m!\ n!\ s!}\  \times \\[3pt]
    & \left[ \left( \frac{m+n-s}{2}\right)!\ \left(\frac{s+n-m}{2}\right)!\ \left(\frac{s+m-n}{2}\right)! \right]^{-1},
    \end{split}
\end{equation}
when $m+n+s$ is even, $m+n\geq s$, $s+n\geq m$, and $s+m\geq n$, or zero otherwise, and the conventional convolution approximation~\cite{Jackson1962}, allows us to define a generalized convolution approximation given by
\begin{equation}
\begin{split}
    &\avg{\rho^{*}_{l} (\mv{k}) \rho_{l_{1}}(\mv{q}_{1}) \rho_{l_{2}}(\mv{q}_{2})} \approx \frac{1}{\sqrt{N}} \delta_{\mv{k},\mv{q}_{1}+\mv{q}_{2}} b_{m,m_{1},m_{2}} \\[3pt]
    & \hspace{2.0cm} \times b_{n,n_{1},n_{2}} S_{ll}(q) S_{l_{1}l_{1}}(q_{1}) S_{l_{2}l_{2}}(q_{2}).
\end{split}
\end{equation}
Here, we have introduced the geometric factor
\begin{equation}
\begin{split}
    & b_{n,n_{1},n_{2}}=(-1)^{-(n-n_{1}-n_{2})/2}\sqrt{n_{1}!\ n_{2}!\ n!} \\[3pt]
    & \left[ \left( \frac{n_{1}+n_{2}-n}{2}\right)!\ \left(\frac{n+n_{2}-n_{1}}{2}\right)!\ \left(\frac{n+n_{1}-n_{2}}{2}\right)! \right]^{-1},
\end{split}
\end{equation}
when $m+n+s$ is even, $m+n\geq s$, $s+n\geq m$, and $s+m\geq n$, or zero otherwise.
Using the generalized convolution approximation we have for the passive contribution of the left vertex
\begin{equation}
\begin{split}
  & \avg{\rho^{*}_{l} (\mv{k}) \Omega^{\dagger}_{\mathrm{eq}} \mathcal{Q} \rho_{l_{1}}(\mv{q}_{1}) \rho_{l_{2}}(\mv{q}_{2})} S^{-1}_{l_{1};l_{1}}(q_{1})  S^{-1}_{l_{2};l_{2}}(q_{2}) = \frac{\rho D}{\sqrt{N}} \\[3pt]
  & \times \delta_{\mv{k},\mv{q}_{1}+\mv{q}_{2}} \bigg( \mv{k}\cdot \mv{q}_{1}\ \delta_{l_{1}0}\delta_{ll_{2}} c(q_{1}) + \mv{k}\cdot \mv{q}_{2}\ \delta_{l_{2}0}\delta_{ll_{1}} c(q_{2}) \bigg),
\end{split}
\end{equation}
where $c(q)=\rho^{-1}[1-S^{-1}(k)]$ depicts the direct correlation function. Note that the passive contribution is thus a straightforward generalization of the standard MCT vertex. Furthermore, the passive contribution to the right vertex can be shown to take on an identical form.

For the active contribution to the left vertex, i.e.
\begin{equation}
    \avg{\rho^{*}_{l} (\mv{k}) \delta\Omega_{\mathrm{T}}^{\dagger} \mathcal{Q} \rho_{l_{1}}(\mv{q}_{1}) \rho_{l_{2}}(\mv{q}_{2})} S^{-1}_{l_{1};l_{1}}(q_{1})  S^{-1}_{l_{2};l_{2}}(q_{2}),
\end{equation}
we recall that $\mathcal{Q}=\mathcal{I}-\mathcal{P}$ and first consider the term
\begin{equation}
    -\avg{\rho^{*}_{l} (\mv{k}) \delta\Omega_{\mathrm{T}}^{\dagger} \mathcal{P} \rho_{l_{1}}(\mv{q}_{1}) \rho_{l_{2}}(\mv{q}_{2})} S^{-1}_{l_{1};l_{1}}(q_{1})  S^{-1}_{l_{2};l_{2}}(q_{2}).
\end{equation}
Using the generalized convolution approximation and~\cref{eq_omega}, this term can be written as
\begin{equation}
    -  \frac{\zeta^{-1}}{\sqrt{N}}\sqrt{D_{f}\tau_{p}}\ \delta_{\mv{k},\mv{q}_{1}+\mv{q}_{2}} S_{l;l}(k)\ \mv{k} \cdot \left( \mv{b}_{l^{-}l_{1}l_{2}} - \mv{b}_{l^{+}l_{1}l_{2}} \right),
\end{equation}
where we have introduced 
\begin{equation}
\begin{split}
  \gv{b}_{ll_{1}^{\pm}l_{2}}= & \left[  \sqrt{m_{1}+\frac{1}{2}\pm \frac{1}{2}}\ b_{m,m_{1}\pm1, m_{2}} b_{n,n_{1},n_{2}}, \right. \\
  & \left. \sqrt{n_{1}+\frac{1}{2}\pm \frac{1}{2}}\ b_{m,m_{1}, m_{2}} b_{n,n_{1}\pm1,n_{2}} \right].
\end{split}
\end{equation}
Next, we also require an expression for
\begin{equation}
    \avg{\rho^{*}_{l} (\mv{k}) \delta\Omega_{\mathrm{T}}^{\dagger} \rho_{l_{1}}(\mv{q}_{1}) \rho_{l_{2}}(\mv{q}_{2})} S^{-1}_{l_{1};l_{1}}(q_{1})  S^{-1}_{l_{2};l_{2}}(q_{2}).
\end{equation}
Exploiting the relation $2x H_{n}(x)=\sqrt{2(n+1)}H_{n+1}(x)+\sqrt{2n}H_{n-1}(x)$ and using the generalized convolution approximation, the above term can be calculated to give
\begin{equation}
\begin{split}
    & - \frac{\zeta^{-1}}{\sqrt{N}}\sqrt{D_{f}\tau_{p}} S_{l;l}(k) \delta_{\mv{k},\mv{q}_{1}+\mv{q}_{2}} \Big[
     \mv{q}_{1} \cdot \left( \mv{S}_{ll_{1}^{-}l_{2}}(q_{1}) - \mv{b}_{ll_{1}^{+}l_{2}} \right) \\[3pt]
    & \hspace{1.0cm} S^{-1}_{l_{1};l_{1}}(q_{1}) + \mv{q}_{2} \cdot \left( \mv{S}_{ll_{2}^{-}l_{1}}(q_{2}) - \mv{b}_{ll_{2}^{+}l_{1}} \right)S^{-1}_{l_{2};l_{2}}(q_{2}) \Big],
\end{split}
\end{equation}
and is written in terms of
\begin{equation}
\begin{split}
  \gv{S}_{ll_{1}^{\pm}l_{2}}(q)=& \Bigg[  \sqrt{m_{1}+\frac{1}{2}\pm \frac{1}{2}}\ b_{m,m_{1}\pm1, m_{2}} b_{n,n_{1},n_{2}} \\[3pt]
  & \hspace{-0.8cm} S_{m_{1}\pm1n_{1};m_{1}\pm1n_{1}}(q),\ \sqrt{n_{1}+\frac{1}{2}\pm \frac{1}{2}} \\[3pt]
  & \hspace{-0.8cm} b_{n,n_{1}\pm1,n_{2}}b_{m,m_{1},m_{2}}  S_{m_{1}n_{1}\pm1;m_{1}n_{1}\pm1}(q)  \Bigg].
\end{split}
\end{equation}

The only term left to calculate is the active contribution to the right vertex. However, this term can be shown to yield a value of zero and thus does not contribute to the vertices. Combining all results we then have for the memory kernel
\begin{equation}
\begin{split}
    M_{l ; l^{\prime}}(\mv{k},t) & \approx \ \frac{\rho^{2}}{4N}\sum_{\mv{q} \mv{q}^{\prime}} \sum_{l_{1}\hdots l_{4}} V_{ll_{1}l_{2}}(\mv{k},\mv{q},\mv{k}-\mv{q}) \\[3pt]
    & \hspace{-1.5cm} \times \avg{\rho^{*}_{l_{1}}(\mv{q})\rho^{*}_{l_{2}}(\mv{k}-\mv{q}) e^{\mathcal{Q}\Omega^{\dagger}\mathcal{Q}^{\prime}\mathcal{Q}t} \rho_{l_{3}}(\mv{q}^{\prime}) \rho_{l_{4}}(\mv{k}-\mv{q}^{\prime}) } \\[3pt]
    & \hspace{2.0cm} \times V^{\mathrm{eq}}_{l^{\prime}l_{3}l_{4}}(\mv{k},\mv{q}^{\prime},\mv{k}-\mv{q}^{\prime}).
\end{split}
\end{equation}
with the vertices given by
\begin{equation}
\begin{split}
    V^{\mathrm{eq}}_{ll_{1}l_{2}}(\mv{k},\mv{q},\mv{k}-\mv{q})= & \ D \left[ \mv{k}\cdot \mv{q} \ \delta_{l_{1}0}\delta_{ll_{2}}\ c(q) \right. \\[3pt]
    & \hspace{-1.7cm} \left. +\ \mv{k}\cdot (\mv{k}-\mv{q})\ \delta_{l_{2}0}\delta_{ll_{1}}\  c(\abs{\mv{k}-\mv{q}}) \right],
\end{split}
\end{equation}
and
\begin{equation}
\begin{split}
    & V_{ll_{1}l_{2}}(\mv{k},\mv{q},\mv{k}-\mv{q}) = V^{\mathrm{eq}}_{ll_{1}l_{2}}(\mv{k},\mv{q},\mv{k}-\mv{q}) - \frac{\zeta^{-1}\sqrt{D_{f}\tau_{p}}}{\rho} S_{l;l}(k) \\[3pt]
    & \Big[ \mv{k} \cdot \left( \mv{b}_{l^{-}l_{1}l_{2}} - \mv{b}_{l^{+}l_{1}l_{2}} \right)
     + \mv{q} \cdot \left( \mv{S}_{ll_{1}^{-}l_{2}}(q) -\mv{b}_{ll_{1}^{+}l_{2}} \right)S^{-1}_{l_{1};l_{1}}(q) \\[3pt]
    & + (\mv{k}-\mv{q}) \cdot \left( \mv{S}_{ll_{2}^{-}l_{1}}(\abs{\mv{k}-\mv{q}}) -\mv{b}_{ll_{2}^{+}l_{1}} \right)S^{-1}_{l_{2};l_{2}}(\abs{\mv{k}-\mv{q}}) \Big].
\end{split}
\end{equation}
To further simplify the expression of the memory kernel, we employ the MCT-approximation and replace the four-point correlation function with projected dynamics by a product of two-point density correlation functions with full dynamics. This yields
\begin{equation}
\begin{split}
    & \avg{\rho^{*}_{l_{1}}(\mv{q})\rho^{*}_{l_{2}}(\mv{k}-\mv{q}) e^{\mathcal{Q}\Omega^{\dagger}\mathcal{Q}^{\prime}\mathcal{Q}t} \rho_{l_{3}}(\mv{q}^{\prime}) \rho_{l_{4}}(\mv{k}-\mv{q}^{\prime}) } \approx S_{l_{1};l_{3}}(\mv{q},t) \\[3pt]
    & S_{l_{2};l_{4}}(\mv{k}-\mv{q},t)\ \delta_{\mv{q},\mv{q}^{\prime}} + S_{l_{1};l_{4}}(\mv{q},t)S_{l_{2};l_{3}}(\mv{k}-\mv{q},t)\ \delta_{\mv{k}-\mv{q},\mv{q}^{\prime}}
\end{split}
\end{equation}
After taking the thermodynamic limit, one finally arrives at 
\begin{equation}
\begin{split}
    M_{l ; l^{\prime}}(\mv{k},t)\approx & \ \frac{\rho}{2} \int \frac{d\mv{q}}{(2\pi)^{2}} \sum_{l_{1}l_{2}}\sum_{l_{3}l_{4}} V_{ll_{1}l_{2}}(\mv{k},\mv{q},\mv{k}-\mv{q}) \\[3pt]
    & \hspace{-1.5cm} \times S_{l_{1} ; l_{3}}(\mv{q},t) S_{l_{2} ; l_{4}}(\mv{k}-\mv{q},t) V^{\mathrm{eq}}_{l^{\prime}l_{3}l_{4}}(\mv{k},\mv{q},\mv{k}-\mv{q}),
\end{split}
\end{equation}
which, using only the equilibrium static structure factor $S(k)$ as initial boundary condition, allows us to self-consistently find a solution for $S_{l,l^{\prime}}(\mv{k},t)$ and in particular for the intermediate scattering function $S_{00,00}(\mv{k},t) \equiv F(k,t)$. We conclude by mentioning that the above derivation can also be straightforwardly extended to three dimensions.


\section{Rotational Symmetry}
\label{AppB}
\noindent Due to the inclusion of the active degrees of freedom, dynamic correlation functions depend explicitly on the direction of the wavevector $\mv{k}$. However, we can bypass this problem by invoking the rotational symmetry of our system to align every wavevector entering correlation functions along a chosen direction. Suppose we rotate our coordinate axes clockwise over an angle $\theta$ (or all particles counter clockwise) such that
\begin{equation*}
    \mv{r}_{j} \rightarrow \mv{r}_{j}^{\prime} = \mv{D}(\theta)\cdot \mv{r}_{j}, \quad \mv{f}_{j} \rightarrow \mv{f}_{j}^{\prime} = \mv{D}(\theta)\cdot \mv{f}_{j}, 
\end{equation*}
with the rotation matrix given by
\begin{equation*}
    \mv{D}(\theta) = \begin{pmatrix}
\cos(\theta) & -\sin(\theta) \\
\sin(\theta) & \cos(\theta)
\end{pmatrix}.
\end{equation*}
As a result the AOUP density mode transforms like
\begin{equation*}
    \rho_{mn}(\mv{k})\rightarrow \frac{1}{\sqrt{N}} i^{m+n}\sum_{j=1}^{N} e^{i\mv{k}^{\prime}\cdot \mv{r}_{j}}H_{m}(\bar{f}^{\prime}_{j,x})H_{n}(\bar{f}^{\prime}_{j,y}),
\end{equation*}
where $\mv{k}^{\prime}=\mv{D}^{\mathrm{T}}(\theta)\cdot \mv{k}$ depicts the rotated wavevector. Realising that $\Omega^{\dagger}$, $P_{\mathrm{eq}}(\Gamma_{\mathrm{T}})$, and $P(\Gamma_{\mathrm{R}})$ are invariant under such a rotation, and rewriting $H_{l}(\bar{f}^{\prime}_{j,x})H_{m}(\bar{f}^{\prime}_{j,y})$ back in terms of $\bar{\mv{f}}_{j}$ allows us to transform correlation functions with wavevector $\mv{k}$ to ones with wavevector $\mv{k}^{\prime}$. Note that $\mv{k}$ is thus rotated clockwise. In the main text we can therefore restrict our discussion to wavevectors aligned along a specific direction, which we have chosen to be the $x$-axis.


\bibliographystyle{apsrev4-1}
\bibliography{all}


\newpage






\end{document}